\documentclass{emulateapj}

\slugcomment{Tentatively scheduled for ApJ 20 October 2004, v614}

\shorttitle{The CM Effect in Early-Type Cluster Galaxies}
\shortauthors{L\'opez-Cruz, Barkhouse, and Yee}
\begin{document}
\title{The Color-Magnitude Effect in Early-Type Cluster Galaxies}
\author{Omar L\'{o}pez-Cruz\altaffilmark{1,4,5}, Wayne A. 
Barkhouse\altaffilmark{2,3,4} and H. K. C. Yee\altaffilmark{3,4}}
\altaffiltext{1}{Instituto Nacional de Astrof{\'\i}sica, Optica y 
Electr{\'o}nica (INAOE), Tonantzintla, Pue. 72840
, M{\'e}xico; omarlx@inaoep.mx}
\altaffiltext{2}{Harvard-Smithsonian, CfA, 60 Garden Street, Cambridge, 
MA 02138; wbark@head-cfa.harvard.edu}
\altaffiltext{3}{Department of Astronomy and Astrophysics, University of 
Toronto, Ontario, M5S 3H8, Canada; hyee@astro.utoronto.ca}
\altaffiltext{4}{Visiting Astronomer, Kitt Peak National Observatory.
KPNO is operated by AURA, Inc.\ under contract to the National Science
Foundation.}
\altaffiltext{5}{Visiting Researcher, Departamento de Astronom{\'\i}a,
Universidad de Guanajuato, A. Postal 144, Guanajuato, Gto., M\'exico}
\begin{abstract}

We present the analysis of the color-magnitude relation (CMR) for a
sample of 57 X-ray detected Abell clusters within the redshift
interval $0.02\leq z\leq 0.18$. We use the $B-R$ vs $R$
color-magnitude plane to establish that the CMR is present in all our
low-redshift clusters and can be parameterized by a single straight
line. We find that the CMRs for this large cluster sample of different
richness and cluster types are consistent with having universal
properties. The k-corrected color of the individual CMRs in the sample
at a fixed absolute magnitude have a small intrinsic dispersion of
$\sim 0.05$ mag.  The slope of the CMR is consistent with being the
same for all clusters, with the variations entirely accountable by
filter band shifting effects.  We determine the mean of the dispersion
of the 57 CMRs to be 0.074 mag, with a small rms scatter of 0.026
mag. However, a modest amount of the dispersion arises from
photometric measurement errors and possible background cluster
superpositions; and the derived mean dispersion is an upper
limit. Models which explain the CMR in terms of metallicity and
passive evolution can naturally reproduce the observed behavior of the
CMR in this paper. The observed properties of the CMR are consistent
with models in which the last episode of significant star formation in
cluster early-type galaxies occurred significantly more than $\sim 3$
Gyr ago, and that the core set of early-type galaxies in clusters were
formed more than 7 Gyr ago.

The universality of the CMR provides us with an important tool for
cluster detection and redshift estimation. A very accurate photometric
cluster redshift estimator can be devised based on the apparent
color-shift of the CMR due to redshift. This calibrator has the
additional advantage of being very efficient since only two bands are
needed. An empirically calibrated redshift estimator based on the
$B-R$ color of the CMR for clusters with $z<0.20$ produces an accuracy
of $\Delta z\sim 0.010$. Background clusters, typically at $z\sim
0.25$ and previously unknown, are found in this survey in the
color-magnitude diagrams as secondary CMRs to the red of the target
cluster CMRs. We also find clear cases of apparent X-ray substructure
which are due to these cluster superpositions. This suggests that
X-ray observations of clusters are also subject to a significant
amount of projection contaminations.

\end{abstract}

\section{Introduction}

Early-type galaxies are the dominant population in low-redshift galaxy
clusters \citep[e.g.,][]{Dre84,KD89}. When compared with other galaxy
types, early-type galaxies can be labeled as a homogeneous class in
terms of the dynamical structure, stellar populations, and the gas and
dust content. The remarkable uniformity of the properties of
early-type galaxies in clusters and the tightness of their fundamental
plane and its projections suggest that these galaxies are coeval, had
their last episode of strong star formation at high redshift
($z~\geq~2$), and have evolved passively since then
\citep[e.g.,][]{B92b,AS93,PDD96,vanD98,SED98,Gladd98,LC02,blake03}. As
larger samples of early-type galaxies become available, the idea that
giant spheroidal galaxies had their major starburst at high redshifts,
regardless of the environment, becomes more popular
\citep[e.g.,][]{Pee02}.

One of the properties of early-type galaxies that can be studied most
straightforwardly is the color-magnitude relation (CMR). This effect
was discovered by \citet{Ba59} who noted that field elliptical colors
become redder as the galaxies become brighter. \citeauthor{Ba59}'s study also
revealed that the average colors of ellipticals are much redder than
those of globular clusters, and concluded, contrary to \citet{Ba44},
that elliptical galaxies were dominated by old Population I stars.  To
arrive at this conclusion, Baum generated a simple population
synthesis model that was based on the, then novel, idea of progressive
metal enrichment of a population by successive generations of stars
\citep{FG56,St56}. Hence, elliptical galaxies may have experienced an
extended period of very efficient star formation during the early
stages of their formation. This idea is now accepted, but still
debatable in terms of our inability to disentangle the effects of age
and metallicity in present day ellipticals
\citep[e.g.,][]{WTF95,FCS99}.

A simple straight line fit can describe the CMR for elliptical
galaxies in an interval of about 8 magnitudes in Virgo
\citep[e.g.,][]{Sa72} and Coma \citep[e.g.,][]{TG93,LC97,SHP98}. This
effect is remarkable because it covers a range in luminosity of about
$10^3$, suggesting that within this range galaxies have shared a
similar evolutionary process.  However, this is not the only way to
get such an effect. For instance, the CMR can be reproduced by
demanding younger galaxies to be more metal rich than older galaxies
of similar luminosity \citep{FCS99}. The potential of the CMR
as a tool for cosmological studies was recognized by \citet{Sa72} and
was characterized for different galaxy environments
\citep{VS77,SV78a,SV78b}. The CMR's application as a distance
indicator has been suggested by \citet{VS77}. The dispersion in the
CMR, which is very small, has been quantified for Virgo and Coma by
\citet[][hereafter BLE]{B92b} and for high-redshift clusters
\citep[e.g.,][]{ESD97,vanD98,SED98, Gladd98,blake03}. These studies
have used the dispersion to constrain the epoch of galaxy formation
and the way they have evolved.

The CMR was modeled as a metallicity effect by \citet{AY87}. The models 
of \citeauthor{AY87} incorporated the ideas of \citet{La74a,La74b} in 
which early-type galaxies are formed by dissipative monolithic collapse and 
the star formation regulated by supernova driven winds. As in \citet{Bow98}, 
we will call Larson's  model the classical model. More recently, 
\citet{Ko97} and \citet[][hereafter KA97]{KA97} have revised the 
models of \citet{AY87} and have included improved libraries for stellar 
evolution, incorporating the effects of the variation of metallicity with 
galaxy mass. Alternatively, the semi-analytical modeling of \cite{KC98}, 
which incorporate the hierarchical galaxy formation scenario, can also 
reproduce the CMR for early-type galaxies by imposing a very early epoch 
for the major mergers. Much effort has been devoted to improving the models 
and the observations; however, it is still difficult to decide whether 
early-type galaxies formed according to the hierarchical or the classical 
model \citep[and references therein]{Bow98}.
 
If cluster early-type galaxies were formed at high redshift, are
coeval and evolving passively, then it should be expected that the CMR
is universal at the current epoch. Universality should be defined in
the present context as the validity for all clusters of the same
rest-frame parameters: slope, intercept and dispersion. The works of
Sandage \& Visvanathan\citep{VS77,SV78a,SV78b} and BLE made a case for
the universality of the CMR for early-type galaxies, albeit based on a
few local clusters. There are other studies that appear to agree with
these results \citep[e.g.,][and references therein]{BO84,DGP89,
MGP94,GB96}. However, some apparent cluster-to-cluster variations of
the CMR have also been reported, but the evidence is not
compelling. For example, \citet{Aar81} reported that the $U-V$ and
$V-K$ colors for galaxies in Coma were bluer than those for galaxies
in Virgo; this result was not confirmed by BLE. Drastic changes in the
CMR's slope for galaxies within the same cluster have been reported
\citep[and references therein]{MGP94}. The model of galactic
cannibalism proposed by \citet{HO78} hinted at an explanation for this
phenomenon.  However, most of the studies at different redshifts, even
as high as $z\approx 0.5$ \citep{ESD97} or higher
\citep{SED98,Gladd98}, have shown little indications of such an
effect. The studies mentioned above utilized relatively small cluster
samples and accurate photometry was limited mainly to the few nearest
clusters. In this paper we show that the CMR in a 57-cluster sample at
low redshifts, with different richness, cluster types and X-ray
luminosities, are consistent with being universal.

This paper is the second of a series resulting from a large multicolor
imaging survey of low-redshift Abell clusters. The paper is organized
as follows. In \S 2 we present the sample selection criteria. In \S 3
we briefly describe the observations and photometric reduction
procedures. Section 4 describes the CMR parameterization and the error
estimate techniques; while \S 5 examines the dispersion of the CMR. We
present a discussion of the results, in particular regarding the
universality and the evolutionary models, in \S 6. In \S 7 we
introduce some observational applications of the CMR based on the
results of this study, including using the CMR as a redshift indicator
and a cluster finding tool. Finally in \S 8 we summarize the
conclusions. More details regarding sample selection, observations,
image preprocessing, catalogs and finding charts can be found in
\citet{LC97}, \citet{bark03}, and Barkhouse, L\'opez-Cruz, \& Yee
(2004a, in preparation, Paper I of this series). A detailed discussion
of the luminosity function of cluster galaxies using this survey can
be found in Paper III (Barkhouse, Yee, \& L\'opez-Cruz 2004b, in
preparation).  Paper IV describes the characteristics of various
color-selected cluster galaxies (Barkhouse, Yee, \& L\'opez-Cruz
2004c, in preparation). Recent cosmological observations indicate that
the best world model agrees with a flat $\Lambda$-dominated universe
that is characterized by $\Omega_{\rm m} \simeq 0.3$,
$\Omega_{\Lambda} \simeq 0.7$ and $H_{\circ}\simeq 70$ km s$^{-1}$
Mpc$^{-1}$ \citep[e.g.,][]{Sp03}.  However, considering that the
effects of dark energy and curvature are negligible at low redshifts
($z < 0.2$) and in order to allow direct comparisons with previous
studies; we, then, for convenience have set, unless otherwise
indicated, $H_{\circ}=
50h_{50}~\mbox{km}~\mbox{s}^{-1}~\mbox{Mpc}^{-1}$ and $q_0=0$
throughout this paper.

\section{The Sample and Observations}

The sample contains Abell clusters selected mainly from a 
compilation of bright X-ray clusters from {\em Einstein}'s IPC made 
by \cite{JF99}. The selection of clusters identified by independent 
techniques (i.e., optical and X-ray) reduces the probability of 
confusing real, gravitationally bound clusters, with apparent galaxy 
over-densities due to projection effects. The initial sample was 
defined under the following selection criteria: 1) the clusters should 
be at high galactic latitude, $|b|\geq 30\arcdeg$; 2) their redshifts 
should lie within the range $0.04 \leq z \leq 0.20$; 3) the Abell 
richness class (ARC) should, preferably, be greater than 0; and 4) the 
declination $\delta\geq 20\arcdeg$. We attempt to follow these 
criteria strictly; however, some ARC=0 clusters were included due to 
observational constraints, such as the lack of suitable clusters at 
certain right ascensions during the observations. The ARC=0 clusters 
were not selected at random, but on the appearance of the X-ray emission 
and the Bautz-Morgan type that indicate a higher richness class. See 
\citet{YLC99} for a discussion of the richness of the cluster 
sample.

This sample includes 47 clusters of galaxies observed in Kron-Cousins 
$B$, $R$, and $I$ at KPNO with the 0.9m telescope using the 
$2048\times 2048$ pixels T2KA CCD \citep[][hereafter the LOCOS 
sample]{LC97,YLC99,LC01}. In addition, two clusters imaged in $B$ and 
$R$ were included from Brown (1997) using the same instrumental setup 
and satisfying our selection criteria. We note that only the $B$ and 
$R$ data are considered in this paper. The field covered by this 
combination of a small telescope plus a 2K CCD is $23.2\arcmin\times 
23.2\arcmin$ with a scale of $0.68\arcsec/\mbox{pixel}$, i.e., $\sim 
1.5h_{50}^{-1}~\mbox{Mpc}$ at $z=0.04$ and $\sim 
6h_{50}^{-1}\mbox{Mpc}$ at $z=0.2$.

A sub-sample of eight clusters from \cite{bark03} are included to 
complement our 49 cluster sample by covering the low redshift interval 
from $0.02\leq z \leq 0.04$. The data were obtained at KPNO with the 
0.9m telescope using the 8k Mosaic camera ($8192\times 8192$ 
pixels). This telescope/detector combination yields a scale of 
$0.423\arcsec/\mbox{pixel}$, with a field-of-view of one square degree 
($\sim 2h_{50}^{-1}~\mbox{Mpc}$ at $z=0.02$ and $\sim 
4h_{50}^{-1}~\mbox{Mpc}$ at $z=0.04$). The clusters from this sample 
were selected using the previous criteria except that ARC=0 clusters 
were not preferentially excluded, although all clusters in the sample 
are detected in X-rays and have a prominent CMR.

The integration times for our 57 cluster sample varied from 250 to 
9900 seconds, depending on the filter and the redshift of the 
cluster. The photometric calibrations were done using stars from 
\citet{La92}. Control fields are also an integral part 
of this survey. We observed in both $R$ and $B$ filters a total of 6 
control fields chosen at random positions on the sky at least 
$5\arcdeg$ away from the clusters in the sample. These control fields 
were observed using the T2KA CCD and the Mosaic camera to a comparable 
depth and reduced in the same manner as the cluster data. All 
observations included in this study were carried out during 1992--93 
and 1996--98.

Table 1 lists the sample with the cluster redshifts. Coma (A1656) was 
included in the LOCOS sample despite its low redshift ($z = 0.0232$). 
The main reason for its inclusion is the vast amount of data for Coma 
available in the literature, which allows us to make direct comparisons 
of our results with others. (We note that only $R$ images of Coma were 
obtained with the Mosaic camera and hence are not considered here.)

\section{Photometric Reductions}

The preprocessing of the images was done using IRAF. The photometric 
reduction was carried out using the program {\sf PPP}\citep[Picture 
Processing Package,][]{Yee91}, which includes algorithms for performing 
automatic object finding, star/galaxy classification and total magnitude 
determination. We also exploited a series of improvements to {\sf PPP} 
described in \citet{YEC96} which decrease the detection of false objects 
and allow star/galaxy classification in images with variable point-spread 
function (PSF). We describe the reduction procedures briefly below; full 
details are given in Paper I.

\subsection{Photometry}

The object list for each cluster is compiled from the {\em R} frames. 
The {\em R} frames are chosen because they are deeper than the images 
from the other filters. On average, about 3000 objects are detected in 
each LOCOS fields  and about 25,000 objects in the mosaic fields. 
Various steps are taken to minimize spurious object detections due to 
bleeding columns, bright star halos, artificial satellite tracks and 
other cosmetic effects.

{\sf PPP} uses the curve of growth of counts to determine {\em total 
magnitudes} of identified objects. A sequence of 30 concentric 
circular apertures is used, starting from a diameter of $\sim 2\arcsec$ 
(three pixels), approximately the diameter of the seeing disk. The 
apertures have a maximum diameter of $20-24\arcsec$, in steps of 
$0.68\arcsec$ or $0.423\arcsec$ (i.e., one pixel). An optimal aperture 
size for each object is determined based on the shape of the growth curve, 
using the criteria described in \citet{Yee91}.

Determining the photometry for galaxies in clusters is more 
complicated than for field galaxies at comparable magnitude 
limits. One of these problems is the large range of galaxy sizes, 
which extends from a few kpcs for dwarf galaxies to $\sim 
1h_{50}^{-1}~\mbox{Mpc}$ for cD galaxies. We find that for bright 
galaxies whose apparent $R$ magnitude is brighter than about 18.5, an 
aperture of $20\arcsec$ is insufficient for the determination of the 
total magnitude. Hence, in a second iteration, the growth curves of 
all objects brighter than $R=18.5$ and classified as galaxies are 
extended to a maximum aperture of $81-120\arcsec$ with a coarser 
radial increment, allowing the total light of even the largest non-cD 
galaxies to be measured at $z\sim0.02$

The presence of close neighbors and crowding is a problem that is more 
acute in cluster galaxy photometry. {\sf PPP} deals with this by 
automatically masking all neighbors within twice the radius of the 
maximum allowable aperture from the object being considered 
\citep[see][]{Yee91}. An additional problem which may adversely affect 
the photometric and color accuracy of cluster galaxies situated near 
the cluster core is the presence of a cD galaxy that, by virtue of its 
large size, may engulf many nearby galaxies in its envelope. To solve 
this problem, the photometry of galaxies near the core is carried out 
after the cD and bright early-type galaxies have been removed using 
profile modeling techniques developed by \citet{Br97}.

\subsection{Color Determination}

The colors for the galaxies are determined using fixed apertures on 
the images of each filter, sampling identical regions of galaxies in 
different filters. Due to the large range of cluster galaxy sizes and 
the sampled redshifts range ($0.02\leq z \leq 0.18$) we adopt the 
following scheme. For bright galaxies $(R\leq 17.5)$ at $z\leq 0.06$, 
a maximum physical aperture of $0.011h^{-1}_{50}~\mbox{Mpc}$ is used. This 
aperture varies in angular size from $19\arcsec$ for $z =0.04$, to 
$6\arcsec$ for $z=0.06$.  For bright objects at larger redshifts the 
maximum allowable aperture is fixed at $6\arcsec$, this corresponds to 
$0.025h^{-1}_{50}$ Mpc at $z=0.18$, the largest cluster redshift in 
the sample. This change in aperture for the objects at redshift larger 
than $z=0.06$ is introduced because the angular size of an 
$0.011h^{-1}_{50}$ Mpc aperture becomes too small at those 
redshifts. Using too small an aperture introduces systematic errors 
due to seeing and variations of the PSF; hence, a minimal color aperture 
of $\sim 3\times$ FWHM is used to avoid these effects (the 
average seeing measured in the cluster images is $\sim 1.5\arcsec$). 
Overall the internal accuracy in the color determinations should be 
about 0.005 magnitudes in $B-R$ for bright objects. For faint objects, 
the physical size is often smaller than $0.011h^{-1}_{50}$ Mpc; hence, 
the smallest optimal aperture from the available filters as determined 
by the growth curve profile is used. The errors for faint objects can 
be as large as 0.5 magnitudes in $B-R$.

The approach of using relatively small aperture sizes for color 
determination increases the accuracy, because only the central region 
of each object is used. This, however, comes at the expense of 
assuming no color gradients in the galaxies. The effects of color 
gradients in $B-R$ are very small for early-type galaxies; color 
gradients of only $\Delta (B-R)/\Delta \log r = -0.09\pm 0.02$ mag per 
dex in radius \citep[e.g.,][]{Pel90,Wu04} are present in elliptical 
galaxies. However, larger color gradients are found in late-type 
galaxies \citep[e.g.,][]{deJ96,gad01}; galaxies could be bluer by as 
much as $B-R=0.8$ mag between the nuclear and the outer regions 
\citep[see][for an alternative viewpoint]{Tay03}. However, those 
color-gradients do not seriously affect our color determinations, 
since these very late-type galaxies are not a significant population 
in rich clusters of galaxies for our redshift range.

We note that the total magnitude of a galaxy is determined using the 
growth curve from the $R$ image, while the total magnitudes in the 
$B$ and $I$ images are determined using the color differences with 
respect to the $R$ images \citep[see][for more details]{Yee91}.

\subsection{Star/Galaxy Classification}

Star-galaxy classification is a very important issue in wide-field 
galaxy photometry, because the foreground stellar contribution relative 
to galaxies is large at the bright end and has large variations from 
field to field. {\sf PPP} uses a classifier that is based on the 
comparison of the growth curve of a given object to that of a reference 
PSF. The reference PSF is generated as the average of the growth curves 
of high signal-to-noise ratio, non-saturated stars within the frame. 
The classifier measures the ``compactness'' of the object by effectively 
comparing the ratio of the fluxes of inner and outer parts of an object 
with respect to the reference PSF.

An additional problem with wide-field imaging is the PSF variation 
across the frame, which, although not severe, is clearly detectable in 
these data, especially for the mosaic images. \citet{YEC96} describe a 
procedure of using local PSFs for star/galaxy classification to 
compensate for this effect. Local PSFs generated from a set of bright 
stars, ranging in number from 30 to 200 per frame and distributed over 
the whole image, are used for classification in these images. The 
detailed procedure, including visual verifications and reliability, is 
discussed in Paper I.

\subsection{Calibration to the Kron-Cousins Standard System}

Instrumental magnitudes are calibrated to the Kron-Cousins system by 
observing standard stars from \citet{La92}. Due to the large 
field, up to 45 standard stars can be accommodated in a single frame. 
The color properties of the standard stars cover a large color range 
that encompasses those of elliptical and spiral galaxies.

We observed standard star fields in each filter $\sim$ three times 
throughout the night. The standard stars are measured using a fixed 
aperture of 30 pixels for the LOCOS frames  and 32 pixels for the mosaic 
data. These aperture sizes are selected as being the most stable after 
measuring the magnitudes using a series of diameters. We adopt the 
average extinction coefficients for KPNO, and fitted for the zero points 
and color terms. The airmass terms for the LOCOS data are held fixed to the 
values $-0.270$, $-0.100$ and $-0.040$ (in units of magnitudes per air 
mass) for $B$, $R$, and $I$, respectively. For the mosaic observations, 
the adopted average extinction coefficients are fixed at $-0.243$ for 
$B$ and $-0.097$ for $R$. Nightly solutions for the remaining coefficients 
are obtained. The rms in the residuals of individual fittings is in the 
range 0.020--0.040 mag, which is comparable to the night-to-night scatter 
in the zero points. This can be considered as the systematic calibration 
uncertainty of the data.

\subsection{Completeness and Final Catalogs}

The final galaxy catalogs are generated using all the information and 
corrections derived in the previous sections. For data obtained under 
non-photometric conditions, single cluster images were obtained during 
photometric nights in order to calibrate the photometry (three 
clusters in total). A final step involves the determination of the 
completeness limit. A fiducial $5\sigma$ detection limit is determined 
for each field by calculating the magnitude of a stellar object with 
brightness equivalent to having a $S/N = 5$ in an aperture of 
$2\arcsec$.  This is done by scaling a bright star in the field to the 
$5\sigma$ level.  However, the $5\sigma$ limit is fainter than the 
peak of the galaxy count curve and hence is below the 100\% 
completeness limit for galaxies. A conservative 100\% completeness 
limit is in general reached at 0.6 to 1.0 magnitudes brighter than the 
$5\sigma$ detection. See \citet{Yee91} for a detailed discussion of 
the completeness limit relative to the $5\sigma$ detection limit.

Each cluster's final catalog contains the object identification 
number, pixel positions, the celestial coordinate offsets with respect 
to the brightest cluster galaxy (BCG), apparent magnitudes $B$, $R$ 
and $I$ ($I$ magnitudes are only available for 47 clusters), and their 
respective errors, object classification, and $R-I$ and $B-R$ 
colors. A more detailed presentation of the catalogs and gray scale 
images of the clusters is given in Paper I, and can be accessed in the 
near future from a Web site.

\subsection{Galactic Absorption Correction}

In order to compare the galaxy colors in a consistent manner, we have 
to correct for the extinction produced by our own galaxy. The corrections 
for the filters used in this study are \citep{PL95}:
\begin{eqnarray}
\nonumber
A_{B} = 4.05 E(B-V), \\
\nonumber
A_{R} = 2.35 E(B-V).
\end{eqnarray}

\noindent
The values of the galactic extinction coefficients can be calculated
from the \citet{BH82} maps, using the reported $E(B-V)$ values, or
directly from the $A_{B}$ tabulations for bright galaxies \citep{BH84}
with coordinates in the vicinity of our pointed observations using
NED. Alternatively, extinction maps based on dust infrared emission
have been provided by \citet{SFD98}. We found that $A_{B}$ values, for
the cluster in our sample, determined by \citeauthor{SFD98} are
systematically larger than those of \citeauthor{BH82}, but in most
cases the difference is not larger than 0.1 mag. In order to
make direct comparison with previous works, we have adopted
\citeauthor{BH82} determinations. We are aware that both schemes have
their own limitations and biases: the difference and uncertainties
between these two extinction schemes are, therefore, considered as a 
systematic error. In Paper I we provide the extinction values used for
each cluster.
 
\subsection{The Color-Magnitude Diagram}

We plot the $B-R$ color versus $R$-band magnitude to create the 
color-magnitude diagrams (CMD). Both the magnitudes and colors have 
been corrected for galactic extinction. The entire set of cluster CMDs 
can be found in Paper I. As examples, Figure 1 shows the CMDs for six 
of the clusters, demonstrating the various properties of the CMR. 
Following the terminology introduced in \citet{MGP94} we can readily 
distinguish three zones: the ``blue'' zone, which is populated by 
late-type galaxies both in the cluster and in the field, and 
early-type galaxies with redshifts lower than that of the cluster; the 
``sequence'', which contains galaxies that have a high probability of 
being early-type cluster members at the cluster redshift 
\citep[e.g.,][]{DGP89,Bi95,TG93,Yee96, Vaz01}; and finally, the 
``red'' zone, which is populated by galaxies which have redshifts 
higher than that of the cluster.

\section{The Parametrization of the CMR}

Different approaches have been applied to quantify the CMR. In most
cases, the analysis has only been applied to objects with confirmed
early-type morphology and membership \citep[e.g., BLE;][]{VS77}. Because the
presence of outliers is minimized by membership and morphology
information, simple least-squares methods can be applied in these
cases. In the present study we do not have redshift information for
the majority of the cluster galaxies nor galaxy morphological
classifications. Nevertheless, as illustrated by the CMDs, the CMR
stands out, and it can be easily traced visually.  Robust methods can
be applied to provide a statistical fit that can be resistant against
the influence of outliers. A few schemes include those used by BLE,
\citet{MGP94}, \citet{ESD97}, \citet{Gladd98}, and
\citet{Pim02}. Below we describe the robust fitting technique used in
this paper.

\subsection{Robust Regression Based on the Biweight Method}

The red sequence of early-type galaxies can be parameterized using a simple 
straight line. Fitting this line to the red sequence of the CMR is not an 
unambiguous procedure. In practice, the presence of a significant number 
of outliers, such as background galaxies, tends to skew the derived solution. 
Simple least-squares methods are ineffective in this situation and more 
``robust'' estimators, such as the Least Absolute Deviation \citep{PT92}, 
have been used \citep{LC97}. After numerous iterations using a variety 
of strategies, a technique utilizing the biweight method \citep{BFG90} 
was adopted. This method is found to be very robust and not greatly 
influenced by the presence of outliers. A variant of this method 
has been implemented by several groups to effectively fit the CMR 
\citep[e.g.,][]{TCB01,Pim02}. 

The procedure used in this study to successfully fit the CMR is 
summarized as follows: An initial cut is made to each cluster catalog 
to cull galaxies further than $0.2\,r_{200}$ from the adopted cluster 
center, which is typically the cD or brightest cluster galaxy. The 
value of $r_{200}$, the radius within which the average density is 200 
times the critical density, is derived photometrically for each 
cluster based on cluster richness \citep[][see paper III for a 
detailed discussion]{YE03}, and is approximately equal to the virial 
radius \citep[e.g.,][]{CL96}. The radius criterion provides a less 
biased measurement of the CMR since our cluster images cover a large 
range in linear size. By using the same dynamical radius for each 
cluster we minimize the influence of using different cluster-centric 
radii, which can potentially affect our CMR measurements 
\citep[e.g.,][]{Pim02}.  The value of $0.2\,r_{200}$ was chosen since 
it is the maximum radius in which all 57 clusters are fully 
observed. For this initial cut of the cluster catalog, galaxies 
fainter than $\sim 3$ magnitudes brighter than the completeness limit 
are removed in order to help reduce the presence of outliers. Once the 
modified cluster catalog is generated, an estimate of the slope and 
intercept is made by visually inspecting the CMR. A set of deviations 
given by $d_{i}=y_{i}-(a+bx)$, where $x_{i}$ and $y_{i}$ are the 
observed values of $R$ and $B-R$ for each galaxy, $b$ is the slope, 
and $a$ is the intercept, is calculated for each galaxy. A biweight 
function is then used to calculate the location and scale. In this 
context, the {\em location} and {\em scale} are statistical quantities 
that describe where the data are mostly found (e.g., median) and the 
spread in the data about this value (e.g., dispersion), 
for the set of deviations (Beers et~al. 1990). This function is given by:
\begin{equation}
f(u) = \left\{ \begin{array}{r@{\quad:\quad}l}u(1-u^{2})^{2}&|u|\le
1\\ 0 &|u| > 1,\end{array}\right.
\end{equation}
where $u=(d_{i}-M)/(c\cdot MAD)$, $M$ is the median, $MAD$ is the 
median absolute deviation from the median, and $c$ is the ``tuning'' 
constant (taken to be 9.0, see Beers et~al. 1990). The value of the slope 
and intercept are than incremented by a small amount $\epsilon$ and a 
new set of deviations constructed. The location and scale are 
determined for this new set, and the process is repeated for a range 
of slope and intercept values that bracket the estimated best-fit 
values. The grid of slope and intercept values are searched to 
determine the best-fit value of $a$ and $b$ that minimizes the 
location and scale of the deviations.

\subsection{Error Estimate and Outlier Rejection}

We apply the non-parametric bootstrap method \citep{ET86,BR93} to 
derive the errors. \cite{BS83} have analytically estimated that $N\sim 
n\log^{2}n$ (where $N$ is the number of resamples to generate and $n$ 
is the number of elements in the original data set) resamples give a 
good approximation of the underlying density distribution. We adopt 
the median values for the slope and intercept derived from $N = 
n\log^{2}n$ bootstrap resamplings as the solution for the fit. These 
values are slightly different from the ones derived from the direct 
fit of the actual data.  The standard errors for the median slope and 
intercept are the rms of these quantities from the bootstrap samples.

In some clusters the standard errors of the slope are large, with the 
uncertainty comparable to the slope itself. We identify three reasons 
for this occurrence: a) the presence of a second CMR produced by a 
serendipitous cluster or group of galaxies at a higher redshift; b) 
when the cluster is poor and there are gaps (in $R$ magnitude) in the 
CMR; and c) for the higher-redshift clusters where the clusters show 
indications of the Butcher-Oemler effect \citep{BO84}. 

For those clusters with large error estimates, a second pass with 
outlier rejection is performed which allows the fitting to move closer 
to the line defined by the CMR and produce more reasonable error 
estimates. Points that are $2\sigma$ away from the main fit are 
rejected and the procedure is repeated. Three cases can be recognized 
in our second pass: a) no effect at all, b) producing a similar fit as 
the first pass but with a smaller error estimate, or c) passing from 
one unique fit to another one. By inspection, we normally recognize 
that the initial fit is already a good fit to the CMR. The second pass 
is introduced to refine the solution and reduce the estimated standard 
errors. A three sigma rejection criterion usually results in case (a). 
Rejection at the one sigma level invariably ends in case (c). The 
optimal approach is obtained using a two sigma rejection procedure 
that results in a new fit consistent with the previous iteration, but 
with smaller computed errors, i.e., case (b). In some cases where the 
CMR is poorly populated, the errors become larger after applying the 
rejection criteria, a consequence of a large change in the fit. We 
note that only three clusters required a second pass with outlier 
rejection. The method outlined above was found to produce an accurate 
characterization of the CMR, as can be seen by visual inspection of 
the color-magnitude diagrams in Figure 1. This procedure was also used 
to successfully fit the CMR of A2271 and A2657 from the sample of 
\citet{LC97}, which could not be achieved by utilizing the Least 
Absolute Deviation \citep{PT92} method used in that study. 

\subsection{Results}

The resultant parameters of the CMR fits are given in columns 3 and 4 
of Table 1, which tabulate the slope and intercept and their uncertainties 
for each cluster. The fit of the CMR is shown in the example CMDs in 
Figure 1 as a dashed line with the fit equation shown. The plot of the 
fits of all the clusters can be found in Paper I.

\section{The Dispersion of the CMR}

The dispersion of the CMR is a useful diagnostic for the star
formation history of early-type galaxies in clusters (e.g., BLE). In
order to provide a robust measure of the dispersion of the CMR, we
create for each cluster (using an identical radius cut as that used in
determining the CMR fit parameters, \S 4.1) a $B-R$ color distribution
which is rectified for the inclination of the CMR using the fit
parameters. The galaxies are binned into 0.1 mag intervals in the
rectified $B-R$ color. An absolute magnitude limit of $-20.2 + 5\log
h_{50}$ (2.0 mag below the mean $M^{*}$ in the LF of the cluster
galaxies; see Paper III) is used to form the distributions.

The error in the color bin counts is assumed to be Poissonian. For 
each cluster a background galaxy color distribution is formed by 
applying the identical rectification and binning to the control 
fields. The background counts in color bins are obtained by averaging 
the control fields, and the uncertainty is generated by calculating 
the standard deviation of the counts from the six control fields in 
each bin. The background is then subtracted bin by bin from the 
cluster color distribution, and the total error per bin is the sum in 
quadrature of the error in the background and the error in the cluster 
color bin counts.

The histogram of the rectified color distribution is shown as an 
insert on the upper left of each CMD in Figure 1. On the CMD the 
$-20.2+ 5\log h_{50}$ mag counting limit is marked by a dotted line, 
while the 100\% completeness magnitude is shown as a solid line. The 
histogram shows the total count distribution, with the white outline 
representing the contribution of the background counts, and the 
background-subtracted net counts in dark shading.

A Gaussian fit is applied to the resultant color distribution to obtain 
a dispersion. This fit is shown as a solid line on the color histogram. 
Since the color distribution also contains galaxies other than 
early-types (as in the case when membership and morphological type 
information is available), blue cluster galaxies and clustered foreground 
and background galaxies contaminate the tails of the distributions. 
Hence, the Gaussian fit is performed by forcing it to fit the peak region 
properly. Visual inspection of the fits indicates that this method 
produces acceptable estimates of the observed width of the CMR.

It is found that the average dispersion of the CMR is 0.074 mag. The 
rms of the distribution of the dispersion is 0.026 magnitudes. We can 
consider this to be an upper bound of the true average dispersion of 
the CMR, since a significant amount of the dispersion must in fact 
arise from errors introduced both by photometric measurement and 
stochastic background galaxy contamination. This is supported by the 
fact that for clusters at lower redshift, where both effects are 
minimized, the CMRs have a lower average dispersion. The mean CMR 
dispersion for the 9 clusters with $z<0.04$ is $0.061\pm 0.029$, where 
the uncertainty is the rms of the distribution; for the remaining 48 
clusters, the mean is $0.076\pm 0.025$. Thus, the mean of the 
low-redshift sub-sample is smaller than the high-redshift sample.

For comparison, a fainter magnitude cutoff of $M_{R}=-17.5+ 5\log 
h_{50}$ is used in fitting a Gaussian to the rectified 
background-subtracted cluster galaxy counts. For a few higher-$z$ 
clusters whose 100\% completeness limit (1.0 mag brighter than the 
$5\sigma$ limit) is brighter than $-17.5+ 5\log h_{50}$, we use the 
completeness limit as the counting limit. The mean CMR dispersion 
obtained using this fainter magnitude limit is found to be 0.106 mag, 
with an rms scatter of 0.040 mag. The larger measured dispersion 
of the CMR using a fainter magnitude cutoff is a reflection of the 
increase in the photometric uncertainty rather than an increase in the 
intrinsic CMR dispersion. This is supported by the fact that the mean 
$B-R$ photometric error using the $-17.5+ 5\log h_{50}$ magnitude 
limit is $0.134\pm 0.164$ mag while for the bright magnitude limit, 
$-20.2+ 5\log h_{50}$, it is $0.040\pm 0.055$ mag (the uncertainties 
are the rms of the distribution).

Estimating the contribution of photometric and background 
contamination errors to the CMR dispersion is a complex matter. This 
is because the error bars are not constant over the magnitude range 
from which we compute the CMR dispersion. In a future paper, we will 
analyze the magnitude of the intrinsic dispersion of the CMR in detail 
using Monte Carlo simulations of the data set based on measured 
photometric error distributions. For the current paper, we will use a 
simple estimate to show that the correlation between the dispersion 
width and redshift is not due to an evolutionary effect, but reflects 
the higher photometric uncertainties of the data at higher 
redshifts. Since our color errors are primarily dominated by the 
errors in the $B$ photometry, we can use the mean $B$ photometry 
errors of the galaxies brighter than the $-20.2+ 5\log h_{50}$ mag 
limit as a proxy for the contribution of photometric error to the CMR 
dispersion.

In Figure 2 we plot the ratio of the CMR dispersion to the average $B$ 
photometric error as a function of redshift of the clusters. From the 
plot it is clear that at $z\ga 0.04$, the mean photometric error is 
$\sim 1/3$ of the dispersion of the CMR; whereas at lower $z$, the 
photometric error is negligible compared to the CMR dispersion. There 
are a number of clusters showing a significantly larger CMR dispersion 
than the mean photometric error. There are 10 clusters with $z\la 
0.04$ which have the dispersion-to-photometric error ratio larger than 
$\sim 30$.  These are A260, A496, A634, A779, A999, A1142, A2152, 
A2247, A2634, and Coma. If we use the quadratic differences of the 
observed dispersion and the average photometric error as a crude 
estimate to the intrinsic dispersion, we find that they have values 
ranging from 0.028 mag (for A634) to as high as 0.126 mag (A2152). We 
note that the cluster with the largest estimated intrinsic dispersion, 
A2152, belongs to the group of clusters which appear to have a 
contaminating CMR from a higher redshift background cluster (see 
Section 7.2). Hence, it appears that both photometric errors and 
possible contaminations from background clusters contribute a 
significant part, if not most, of the apparently large dispersion 
values, and we do not find definitive evidence that there is a large 
range of dispersion in the CMRs.

\subsection{Radial Dependence of the CMR}

The measurement of the CMR is based on a region surrounding each 
cluster of $0.2\,r_{200}$ in radius. In order to look for radial 
changes in the measured properties of the CMR, we have repeated our 
cluster CMR fits using an annulus extending from 0.2--$0.5\,r_{200}$. 
Due to the increase in the extent of the measured region, only 33 
clusters have the necessary areal coverage. A comparison of the same 
33 clusters for the two different annular regions yields a mean CMR 
dispersion of $0.081\pm 0.030$ mag for the inner region and $0.090\pm 
0.034$ mag for the outer radial region. The uncertainties are 
calculated from the rms of the distribution. For the slope of the CMR, 
the 0--$0.2\,r_{200}$ region has a mean value of $-0.052\pm 0.008$, 
while the 0.2--$0.5\,r_{200}$ outer region has a mean slope of 
$-0.053\pm 0.009$ (rms uncertainties). Color changes between these 
two regions can be quantified by comparing the difference in 
$B-R$. Using the CMR fits, the average color difference at $R=17$ 
between the inner and outer radial region is $\Delta (B-R)=0.040\pm 
0.055$, where the uncertainty is the rms. The comparison of the mean 
CMR dispersion, slope, and color difference between the inner and 
outer annuli provides evidence of a slight increase in the CMR 
dispersion and a blueward $B-R$ color change with increasing 
cluster-centric radius. This is probably due to the increasing blue 
fraction with radius \citep[density morphology relation;][]{Dre80} and 
the greater relative number of background galaxies (see Paper IV).
   
\section{Discussion}
\subsection{Elliptical Galaxy Formation and Evolution in Clusters}

The classical understanding of the CMR in terms of metallicity rests 
on the idea that massive stellar systems are able to experience longer 
episodes of efficient star formation. In these systems a large number 
of supernova explosions are necessary to inject enough energy into the 
ISM in order to surpass the binding energy of the galaxy to expel the 
remaining gas and halt star formation activity. It has been shown 
\citep[e.g.,][]{Car84} that the removal of enriched gas by 
supernova-driven winds is more efficient in low-mass systems. Hence, 
giant ellipticals are redder due to a longer episode of subsequent 
star formation that result in the higher metallicity of their stellar 
population; while dwarf ellipticals have had shorter episodes of star 
formation that result in a lower metallicity of their star population, 
and hence their bluer colors. Passive evolution sets in once the star 
formation activity has ceased. Any recent episode of star formation 
will cause variations of the slope and broadening of the dispersion of 
the CMR. A possible extreme situation that one can consider is that 
the CMR can break up if in a given cluster only the giants or the 
dwarfs have experienced some recent episode of star formation.

At the present epoch the color effects due to dissipative galaxy 
formation cannot be distinguished from age effects. However, KA97 have 
shown that, if the redder color of more luminous early-type galaxies 
are due to age, then the CMRs should not exist at high 
redshift. Observational evidence has shown that clusters of galaxies 
with $z$ up to 1.2 have prominent CMRs (see \S 1). Therefore, KA97 
concluded that the origin of the CMR is due to metallicity effects 
induced by supernova-driven winds. The interpretation that early-type 
galaxies in clusters were formed at redshifts $>2$ should produce very 
similar CMRs in current epoch clusters. In the next section, we will 
look at the universality of the properties of the CMR and the 
resultant constraints on the formation and evolution of early-type 
galaxies in clusters. 

\subsection{The Universality of the CMR}
\subsubsection{The Color of the CMR}

We can compare the rest-band colors of the CMR in our sample by 
computing the k-corrected $B-R$ color of the fitted CMR at a fixed 
absolute magnitude. We pick a fiducial absolute magnitude of 
$M_R=-22.2+5\log h_{50}$, the mean $M^{*}$ in the LF of the cluster 
galaxies (Barkhouse, Yee, \& L\'opez-Cruz 2004b, in preparation). The 
$B-R$ color of the red sequence of galaxies in the CMR is computed at 
the apparent magnitude equivalent to the fiducial absolute 
magnitude. This color is then corrected to the rest bands by using 
computed k-corrections for $B$ and $R$ for E/S0's from \citet{CCW}.

For the 57 clusters with a CMR fit, a mean k-corrected $B-R$ color at 
$M_R=M^{*}=-22.2+5\log h_{50}$ of $1.539\pm 0.063$ is obtained, where 
the uncertainty is the rms dispersion of the distribution. A 
conservative estimate of the systematic uncertainties in the 
calibration of the $R$ and $B$ photometry is $\sim$ 0.03 mag for each 
filter, producing a systematic uncertainty of 0.042 mag in 
$B-R$. Hence, the intrinsic dispersion of the color of the CMR in this 
large sample over a small redshift range is about 0.05 mag. This small 
dispersion is very similar to that estimated by Smail et al. (1998), 
who, using 10 clusters at $0.22<z<0.28$, found also a small observed 
dispersion of the color of the CMR of $\sim 0.04$ mag, consistent with 
an intrinsic dispersion of 0.03 mag. Our measured intrinsic scatter is 
also consistent with values measured by \citet{SED98} for a sample of 
19 clusters spanning a redshift range from $0.03$ to $0.9$ (see their 
Fig. 6).

\subsubsection{The Dispersion of the CMR}

The dispersion of the CMR can potentially serve as a statistical
indicator of the star formation history of early-type cluster galaxies
(e.g., BLE). For example, if the dispersion is connected to the
initial formation, it could be used to estimate the spread of the
formation age of the galaxies. Or alternatively, the dispersion could
be an indication of additional episodes of star formation.

Typically, most investigations have found relatively small dispersions 
in the CMRs, although they are all based on very small samples and 
usually sampling only over a range of four or five magnitudes. For 
example, BLE found a very small dispersion of 0.05 mag in 
$U-V$ with about 0.03 attributable to photometric errors. In $B-R$, 
and over a similar baseline of magnitude, we obtain a comparable 
measurement for Coma --- a dispersion of 0.06 mag with a mean 
photometric error of 0.01 mag, using the background subtraction 
technique (with no membership and morphological type 
information). Using HST data with redshift and morphological type 
information, \citet{vanD98} performed very detailed analysis of the 
CMR of the rich EMSS cluster 1358+62 at $z=0.33$, and found again very 
small dispersions of 0.02 to 0.04 mag, depending on the exact galaxy 
sample chosen. \citet{ESD97}, using HST data for the core regions of 
three $z\sim 0.55$ clusters, found dispersions ranging from 0.07 to 
0.11 mag. For high redshift clusters ($z\sim 1.3$), \citet{vanD01} and 
\citet{blake03}, using HST data, measured dispersions of 0.04 and 0.03 
mag, respectively. Our mean dispersion of 0.074 mag can only serve as 
an upper limit to the ensemble average, and is consistent with these 
measurements. Despite our large sample, we have found no definitive 
evidence of clusters with dispersion broader than those reported in 
the literature, i.e., larger than $\sim 0.1$ mag. This lack of broad 
dispersion clusters is consistent with an early formation epoch and 
the lack of significant recent star formation in the early-type 
galaxies in clusters.

\subsubsection{The Slope of the CMR}

Comparing the variation of the slope of the CMR to those predicted by 
models is more robust than comparisons based on absolute color 
evolution. Uncertainties in the transformations to standard 
photometric systems, both in the observations and the models, and 
k-corrections can considerably reduce the significance of any 
photometric test; whereas the slope of the CMR is subject only to 
internal errors. \citet{Gladd98} exploit this robustness of the 
comparison between models and observations of the slope of the CMR 
for clusters at  $z\sim 0$ to 0.8. For the low-redshift anchor, 
they used redshift-binned composites from a sub-sample of the full 
cluster sample CMRs that were available from \citet{LC97}. Comparing 
with CMRs of higher redshift clusters from HST data, \citet{Gladd98}
concluded that the formation epoch, assuming a coeval early-type 
population in clusters, is conservatively larger than $z\sim 2.5$.

The 57 clusters from our sample provides one of the largest single 
samples of CMRs over a small redshift range obtained in a homogeneous 
fashion. This allows us to compare a large sample of clusters of 
various richness and cluster morphological types, at about the same 
epoch, to test the universality of the CMR in a single epoch. In 
Figure 3 we plot the distribution of the $B-R$ CMR slopes of 
individual clusters as a function of the cluster redshift. The slopes 
have a very narrow range of values between $-0.03$ to $-0.08$, with a 
definite trend of steeper values towards larger redshifts.

To interpret the slopes of the CMRs, we appeal to the models of
KA97. The KA97 models are sophisticated stellar population synthesis
models which include new libraries for stellar evolution accounting
for the effects of metallicity. Kodama (1996, private communication)
has repeated the calculations in KA97 for the filters employed by this
study. The transmission curves for the $B$, $R$ and $I$ filters were
obtained from KPNO. An epoch of galaxy formation of 15 Gyr
($z_{f}\approx 5.4$) is assumed. A more critical parameter is the
epoch when the supernova driven winds ($t_{gw}$) are established.
KA97 have calibrated $t_{gw}$ using the observed CMR for Coma (BLE)
based on $U-V$ colors. The resultant variations of the model slope as
a function of redshift is plotted on Figure 3 as a dashed line. We
remark that this model is not a fit, i.e., no freely adjustable
parameter based on our cluster data has been applied. The model was
generated independently using as the anchoring calibration
BLE's data for Coma.  The model clearly provides an excellent
fit, including the slight steepening trend of the slope with
redshift. The very slight apparent slope variations seen here are due
to the CMRs being sampled at slightly bluer rest wave-bands at higher
redshifts, and it is not due to stellar evolutionary effects. Hence,
once the CMR is properly corrected for this band pass effect, we can
conclude that the CMR is very uniform in its parametrization for the
whole sample of 57 clusters.

\subsubsection{The Universality of the CMR and Cluster Evolution}

Our cluster sample contains a diverse set of clusters: it covers a 
richness range over a factor of 4 \citep{YLC99}, from Abell 0 to 
amongst the richest known clusters; it contains clusters that cover 
the full range of the morphological types for schemes in the optical 
and in the X-rays \citep[see][for a review on the morphological 
classification of clusters]{LC03}. The lack of variations in the slope 
of the CMR, along with the very small dispersion in the color, across 
all clusters with such diverse properties allows us to draw the very 
conservative conclusion that no cluster has experienced a significant 
star formation episode in the core during the last 3 Gyr (i.e., the 
look-back time of $z\sim 0.2$). Furthermore, this universality 
indicates that {\it all} current epoch clusters contain a core set of 
early-type galaxies which were formed at a sufficiently distant 
epoch. As demonstrated in \citet{Gladd98}, using models from KA97, it 
can be shown that a significant digression of the CMR slope, in the 
form of a turn-over to a shallower slope, is expected due to the 
differing evolution rates of early-type galaxies with different 
metallicities when the galaxies are at ages younger than $\sim 4$ Gyr 
after formation. Hence, we can conclude that none of these 57 
clusters were formed less than 7 Gyr ago (or $z\sim 0.8$ for 
$H_{0}=70~\mbox{km}~\mbox{s}^{-1}~\mbox{Mpc}^{-1}$, $\Omega_{m}=0.3$, 
and $\Omega_{\Lambda}=0.7$). Given the relatively large sample size, 
one can conclude that it is very unlikely that there are any 
current-epoch clusters (rich enough to be in the Abell sample) which 
were formed less than 7 Gyr ago. We note that \citet{Gladd98}, 
combining 44 clusters from the LOCOS sample and HST data for 6 higher 
redshift clusters, suggested, based on the variation of the CMR slope 
with redshift, that early-type galaxies in clusters form an old 
population with formation redshift $z_{f}\geq 3$. However, because of 
the small number of high-$z$ clusters and their high average richness, 
they could not rule out that {\it some} (likely less massive) clusters 
may form at lower redshifts. Recent studies by \citet{vanD01} and 
\citet{blake03} extend the redshift limit to $z\sim 1.3$ for the 
existence of a well-established early-type cluster population, further 
supporting a formation redshift $z_{f}\geq 3$. A large sample of 
clusters covering a wide range of cluster properties at $z$ between 
0.5 and 1.0 will allow us to place stringent constraints on the 
formation epoch of clusters as a function of cluster properties.

\subsection{The Coma Cluster and the CMR for Dwarf Galaxies}

The Coma cluster (Abell 1656) is the most studied galaxy cluster; its 
proximity and richness makes it ideal for all varieties of cluster 
studies. It was included as an anchoring reference for the properties 
that we measure in the other clusters of our sample. What is striking 
is that, because of the low redshift of the cluster and the depth of 
sampling, its CMR (see Figure 1) can be defined over a range of 8 
magnitudes ($R=12.5$ to 20.5) easily, and it can be detected well into 
our 100\% completeness limit of $R=21.2$ ($\sim -14.2+ 5\log h_{50}$ 
in absolute magnitude). If we define the dwarf galaxy population in 
the same manner as \citet{SHP98}, i.e, galaxies fainter than $R=18$ in 
apparent magnitude (absolute magnitudes fainter than $-18+ 5\log 
h_{50}$), it is apparent that the {\it same} linear fit describes the 
CMR from giants to dwarfs. This result agrees with \citet{Sa72} 
indication that dwarf galaxies follow the CMR defined by bright 
elliptical galaxies in Virgo. Visual inspection of the objects on the 
sequence zone indicates that their morphology resembles dwarf 
elliptical galaxies. Hence, we do not detect any change in slope in 
the CMR over this very large range of galaxy masses. We also note that 
we do not detect any significant changes of slope within the CMR in 
any of our clusters, most of which we can sample to and past absolute 
magnitude $-17+ 5\log h_{50}$ (although, because of the higher 
redshift, the background confusion is usually larger). Thus, if there 
is a bend in the CMR, as has been claimed by a number of studies 
(e.g., Metcalfe et al. 1994), this is not a frequent occurrence at 
these redshifts. The bending may be due to photometric errors or due 
to the working definition of the color aperture.

The slope of the CMR can be used as an age indicator \citep[\S6.2.3 
and][] {Gladd98}; hence, the fact that dwarfs appear to follow the 
same CMR as the giants suggests that dwarf galaxies have a similar 
evolutionary history as giant ellipticals. This seem to be the case: 
\citet{OSR02} using medium-band photometry in the Coma cluster have 
found that dwarf galaxies do follow the CMR defined by giant 
early-type galaxies. However, according to \citeauthor{OSR02} dwarf 
ellipticals, depending on the morphology, could be either younger or 
older than giant ellipticals.

We have explored some observational properties of the CMR. We found 
that these properties can be explained within the classical models. 
However, alternative models can be introduced considering both age and 
metallicity effects \citep[e.g.,][]{FCS99} or alternative formation 
scenarios \citep[e.g.,][]{KC98}. A complete scenario for the formation 
of early-type galaxies in clusters should also incorporate other 
observational constraints, i.e., line strength indices and their 
evolution, and the fundamental plane as its evolution.

\section{Practical Application of the CMR}

The CMR, beside being an important diagnostic in the formation and 
evolution of early-type galaxies, can also serve as a powerful 
observational tool for detecting galaxy clusters and measuring 
photometric redshifts. In the following we briefly examine these 
applications.

\subsection{A Single-Color Redshift Estimator}

In the last few years, there has been a resurgence of interest in and 
applications of photometric techniques for determining redshifts of 
galaxies \citep[see, e.g.,][for a review]{Yee99}. Typically such 
techniques use about four filters, combined with either empirical 
training-set calibrations or model spectral energy distributions, to 
derive the redshift. The CMR of clusters of galaxies, however, offers 
the unique capability of serving as a redshift estimator using only 
two filters, since the population type of the galaxies is known. In 
the observer's frame the CMR of a cluster shifts toward redder colors 
as one progresses in redshift. It is possible to calibrate the changes 
in color of the CMR using our cluster sample as a training set and 
generate a redshift estimator. This indicator, calibrated empirically 
in the observer's frame, will also account for evolutionary effects, 
if any is detectable at these low $z$. In the following we determine 
an empirical correlation between the $B-R$ color of the CMR and redshift.

The zero points from the fits of the CMR (which is equivalent to the 
color of a galaxy at 0 apparent magnitude) in Table 1 show a strong 
correlation with redshift, but the errors are large, as this value is 
greatly correlated with the uncertainty in the slope. A much better 
estimator can be obtained by using a point on the CMR within the 
magnitude range of the observed CMR. We have chosen a fiducial, but 
arbitrary, magnitude of $R=17$. An apparent magnitude is chosen as the 
fiducial because no prior knowledge of the cluster's redshift is 
assumed when determining the redshift of an unknown cluster. The 
$(B-R)_{R=17}$ values are listed in column 6 of Table 1. The average 
error in $(B-R)_{R=17}$ as estimated from the fit to the CMR by 
bootstrap is $\approx 0.016$ mag for clusters at $z\sim 0.04$, and 
$0.030$ mag for clusters at $z\sim 0.15$. In Figure 4 we plot the very 
tight correlation between $(B-R)_{R=17}$ with redshift. The 
correlation between the computed $(B-R)_{R=17}$ and $z$ is fitted with 
a quadratic polynomial anchored on Coma:
\begin{eqnarray}
z & = & 0.0849(+/- 0.0165)(B-R)^2 - 0.1419 (+/- 0.0573)\nonumber \\
 & & (B-R) + 0.0511 (+/- 0.0496).\label{eq:res}
\end{eqnarray}
This fit is shown as a solid line in Figure 4. The dispersion about 
the fit is $\sigma = 0.010$, or a mean $\Delta z/z\sim 0.13$. This 
dispersion is comparable or better than the best $\sigma_z\sim 0.04$ 
that is typically obtained using training set techniques for redshift 
determination using color information from at least four bands for 
$z\sim 0.3$ galaxies \citep[e.g.,][]{Brun97}. The high accuracy of 
this method rest in the fact that we are using a large number of 
galaxies with a known spectral type simultaneously. \citet{PL95} have 
devised a redshift indicator based on the $L_{m}-\alpha$ (the metric 
luminosity and the logarithmic slope of the surface brightness 
profile) relationship for BCGs.  For our sample, we found that the mean 
and standard deviations for $\frac{(z_{obs}-z_{est})}{z_{obs}}$ were 
-0.0164 and 0.180, respectively. These values are comparable to the 
ones found in the \citeauthor{PL95} study; however we have covered a 
redshift range about four times as large.  It is interesting to note that 
the CMD of A2152 (Figure 1) shows the presence of a contaminating 
cluster at $B-R\sim 2$, or $z=0.11$ according to equation 
1. \citet{blake01} has recently reported the discovery of a cluster 
$2.4^{\prime}$ from the center of A2152 at a spectroscopically 
determined redshift of $z=0.13$, and having a red sequence at $B-R\sim 
2.1$. The CMR provides a very reliable redshift estimator that is 
based on a uniform property of elliptical galaxies.

\subsection{The CMR as a Tool for Cluster Detection}

The universality of the CMR, at least at the low-redshift range being 
considered, suggests that the signature of a CMR in a CMD is a good 
indicator of galaxy clusters. In the case of our data, such a 
signature can be used to detect high-density concentrations of 
galaxies in the background or foreground of the target Abell 
clusters. Simple visual inspection of the CMDs (Figure 1, and Paper I) 
shows that the presence of such contaminating clusters is very easily 
seen. Prominent examples include A690, A2152, and A2399. However, a 
more efficient way to detect a background cluster is to use the 
rectified color histograms that are applied to determine the 
dispersion of the CMR (Section 5). A background cluster can be easily 
seen, after background subtraction, as a strong peak with a median 
color redder than the foreground target cluster. By inspection of the 
color histograms, we identify nine clusters (A514, A690, A999, A1291, 
A2152, A2247, A2271, A2399, and A2634) as having a very high 
probability of having a background cluster within their image frames, 
and an additional four (A168, A671, A1795, and A2440) as having a 
possible background cluster superposition. A robust and rigorous 
cluster-finding algorithm based on this idea of the CMR as a cluster 
detection signature, which searches for enhanced density in the 
four-dimensional space of color, magnitude, and x-y sky position has 
been developed by \citet{GY02}, and this algorithm will be applied to 
our cluster data to quantify the background cluster detections in a 
future paper.

At the modest depth of our survey, which allows us to detect clusters 
out to about $z\sim 0.3$, using this crude method of identifying 
clusters from the $B-R$ color distribution, we have detected 9 to 13 
background clusters to our target sample of 57 clusters, or at a rate 
of about 15 to 20\%. The $B-R$ colors of the background clusters are 
typically greater than 2.3 mag, indicating they are at a redshift of 
greater than 0.15 (see Section 7.1). In total, we have 57 cluster 
observations sampling $\sim 15$ square degrees, with an average 
completeness limit of $R=21.2$. Thus we detect $\sim 1$ cluster per 
square degree having a median redshift of $z=0.2$.

There is a general perception that X-ray surveys and observations are 
free from background cluster-cluster contaminations. It is interesting 
to examine the few cases where we find definite optical projections. 
A remarkable case is A514, which has been considered as a standard 
cluster for substructure studies in the X-ray 
\citep[e.g.,][]{Ebel96,JF99,Gov01,Kol01}. The X-ray image of this 
cluster shows many substructures, including a very prominent component 
to the NW \citep{JF99}. The CMD clearly shows a background CMR at the 
color of $B-R\sim 2.8$, corresponding to a redshift of greater than 
0.20, and the galaxies contributing to this background CMR are 
coincident with this large substructure. Thus, at least part of the 
complex X-ray structure of A514 arises from the superposition of 
clusters. Another very clear case of cluster superposition in the 
X-ray is provided by A2152. The X-ray image shows a strong source with 
a NW extension \citep{JF99}. In this case, the CMD clearly shows the 
presence of a background cluster at $B-R\sim 2.0$ and has been 
spectroscopically verified as a $z=0.13$ cluster \citep{blake01}, 
which is coincident with the X-ray extension. Of the remaining seven 
cases of clear background CMRs, only A999 and A1291 do not show 
significant X-ray substructure. Furthermore, of the four clusters 
with possible superposing CMDs, two show significant X-ray 
substructure. Hence, we have found that in a number of instances, 
background cluster-cluster superpositions may well have a significant 
effect on the apparent X-ray morphology of clusters. A more detailed 
analysis of the cluster-cluster superpositions to study the 
correlation between such superpositions and X-ray structures based on 
the more quantitative red-sequence cluster finding algorithm of 
\citet{GY02} will be presented in a future paper.

\section{Conclusion}

In this paper we have explored the properties and applications of the 
CMR using a large sample of 57 low-redshift Abell clusters. We 
demonstrate that simple techniques based on the biweight function 
are able to fit a single line to the CMR. A straight line describes 
the CMR for a range of at least 5 magnitudes for all the clusters in 
the sample, and extending as far as 8 magnitudes when the data are of 
sufficient depth. We find no evidence for a change of slope within the 
CMR over this magnitude range.

The properties of the CMR for this diverse sample of low-redshift 
clusters are remarkably universal. We also find no definitive evidence 
that the dispersion of the CMR is significantly greater than about 0.1 
mag in any cluster. The distribution of the k-corrected color at a 
fixed absolute magnitude of the individual CMRs in the sample have a 
very small dispersion of 0.05 mag. Finally, the slope of the CMR is 
entirely consistent with being identical, once the CMRs are corrected 
to the same rest wave bands. Given the constancy of the parameters of 
the CMR, the universality of this fundamental property of early-type 
galaxies in clusters is established. In agreement with the results of 
\citet{SED98}, we show that our observations are consistent with 
models that explain the origin of the CMR based on galaxy formation 
which produces the color dependence due to varying metallicities. The 
universality of the slope and color of the CMR enable us to place a 
conservative limit that the last episode of significant star formation 
in cluster early-type galaxies happened more than $\sim3$ Gyr ago, and 
that the core set of early-type galaxies in current epoch clusters 
were formed more than $\sim 7$ Gyr ago.

We demonstrate that the CMR provides a powerful tool for the detection 
of clusters and also can be used as a very accurate and efficient 
redshift indicator for groups and clusters of galaxies. The CMR can 
also be effectively used to limit the contribution of background 
galaxies in the statistical determination of cluster membership, in 
that galaxies redder than the CMR can be confidently considered as 
background galaxies \citep{LC97,LC97b,bark03}. We will exploit this 
property in the companion paper on the luminosity function of cluster 
galaxies (Paper III).

\acknowledgments

Many of the results presented in this paper are drawn from OLC's and 
WAB's Ph.D. theses. OLC thanks the hospitality of the following 
centers: High Energy Division of the SAO, the Institute for Advance 
Study of the Norwegian Academy of Sciences, Carnegie Observatories SBS 
(Fulbright Visiting Scholarship, 2001), Caltech's Department of 
Astronomy (FUMEC-AMC Young Researchers Visiting Program, 2002) where 
the results were presented and discussed prior to their 
publication. The authors are grateful to Gus Oemler, George 
Djorgovski, Taddy Kodama, Allan Sandage, Fran\c{c}ois Schweizer, Alan 
Dressler, Richard Bower, Ray Carlberg, Erica Ellingson, Huan Lin, 
Felipe Barrientos, and Bob Abraham for enlightening conversations and 
comments. We thank T.~Kodama for providing the model slopes of the 
CMR, Huan Lin for providing photometric catalogs for five control 
fields, and James Brown for the use of his galaxy profile fitting 
software and photometric data for A496 and A1142.  OLC's research at 
the University of Toronto was sponsored by an overseas scholarship by 
CONACyT-M\'exico, INAOE, The Carl Reinhart Fund and NSERC through 
HKCY's operating grant. OLC's research is supported in part by 
CONACyT-M\'exico Project J32098-E. WAB's research at the University of 
Toronto was supported by The Carl Reinhardt Fund, The Walter C. Sumner 
Fellowship, and NSERC through HKCY's operating grant.  HKCY's research 
is support by an NSERC operating grant.

IRAF is distributed by NOAO, which is operated by AURA, Inc., under 
contract with the NSF. In this paper we have made intensive use of the 
NASA/IPAC Extragalactic Database (NED), which is operated by JPL, 
Caltech, under contract with NASA.


\clearpage

\figcaption{Example color-magnitude diagrams from our cluster sample 
in $B-R$ vs $R$ for the total region covered by our images. The solid 
line shows the 100\% completeness limit, while the dotted line indicates 
absolute magnitude $-20.2$ at the cluster's redshift. The dashed line 
illustrates the best fitting color-magnitude relation for the cluster, 
with the fit equation shown at the lower left. The insert histogram is 
the CMR rectified $B-R$ color distribution for background-subtracted 
galaxies within $0.2\,r_{200}$ of the adopted cluster center. 
The Gaussian fit to the CMR is shown as the thick solid line 
over-plotting the histogram with the dispersion indicated on the upper left 
corner. A total of 6 CMDs are shown. These are: (a) A21, (b) A690, (c) A1213, 
(d) A1656 (Coma), (e) A2152, and (f) A2399. The CMD of the complete 
sample can be found in Paper I. We note that A690, A2152, and A2399 are 
examples of cluster fields which have an apparent background CMR, likely 
arising from background clusters at $z>0.2$. These can be detected both in the 
CMDs and as the prominent second peak in the $B-R$ histograms.}

\figcaption{The ratio of the computed dispersion of the CMR from the 
$B-R$ histogram to that of the mean $B$-band photometric uncertainty 
of the galaxies versus redshift. Note that for clusters with $z>0.04$, 
their CMR dispersion is partly due to photometric errors. For the $z<0.04$ 
clusters, there are a number of clusters which show a significantly larger 
CMR dispersion than expected simply from measurement errors. At least part 
of this dispersion can be attributed to excess background contaminations 
from cluster-cluster superpositions
(see text).}

\figcaption{The variation of the slope of the CMR with redshift. The 
points correspond to the cluster sample. The dashed line is not a fit, 
but the predicted variation derived from a spectral synthesis model 
that accounts for variations due the metallicity and passive evolution 
in elliptical galaxies (KA97). The equation parameterizing this model 
as a function of $z$ is given at the top of the graph.}

\figcaption{The variation of the CMR color at a fixed apparent magnitude 
with redshift. The points correspond to the $B-R$ colors at $R=17.0$ for 
the clusters in this survey as computed from the best fitting CMR. The 
solid line represents a second order polynomial fit to the data 
(see text for details).}

\clearpage

\begin{figure}
\figurenum{1}
\plotone{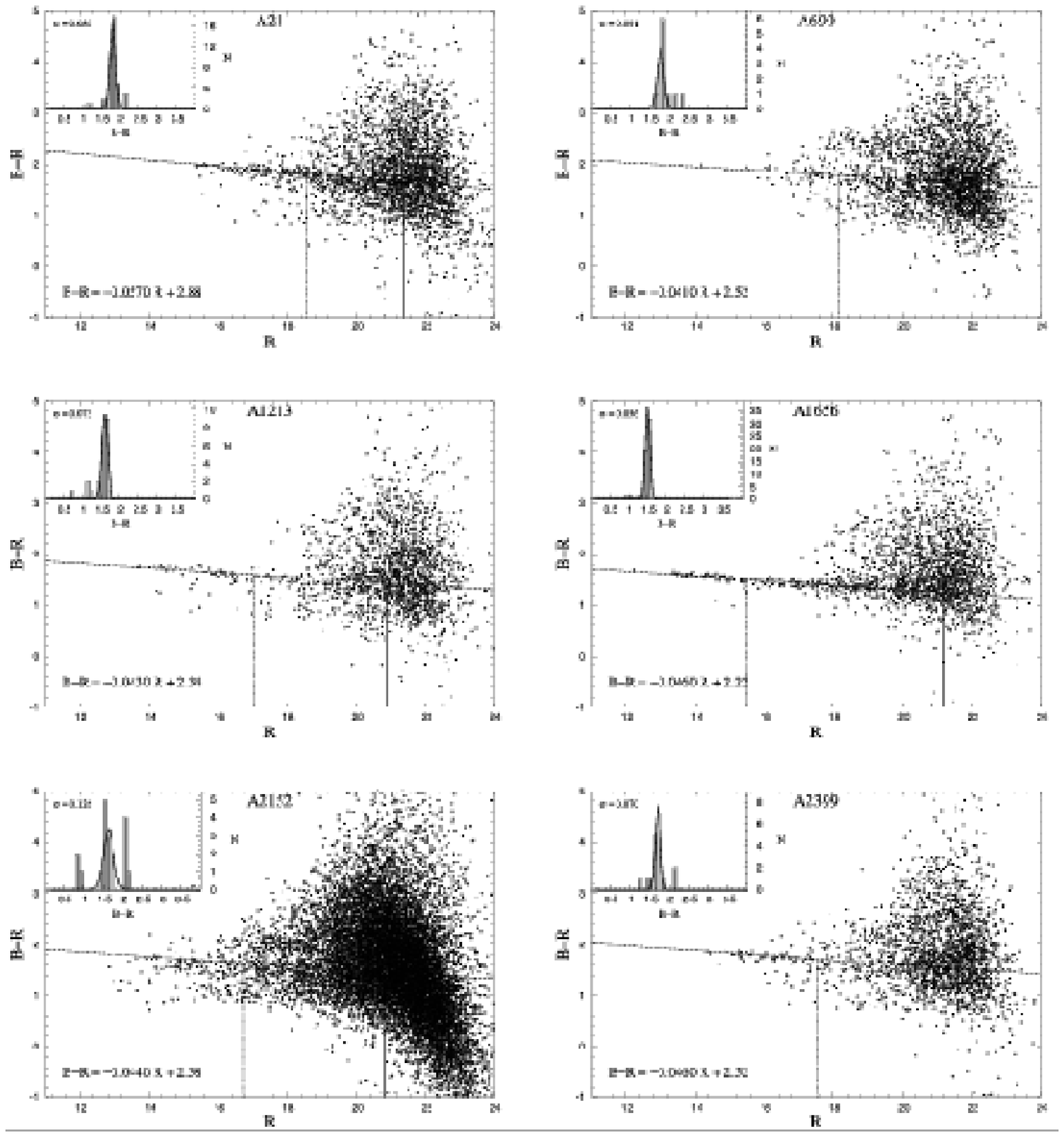}
\caption{}
\end{figure}
\begin{figure}
\figurenum{2}
\plotone{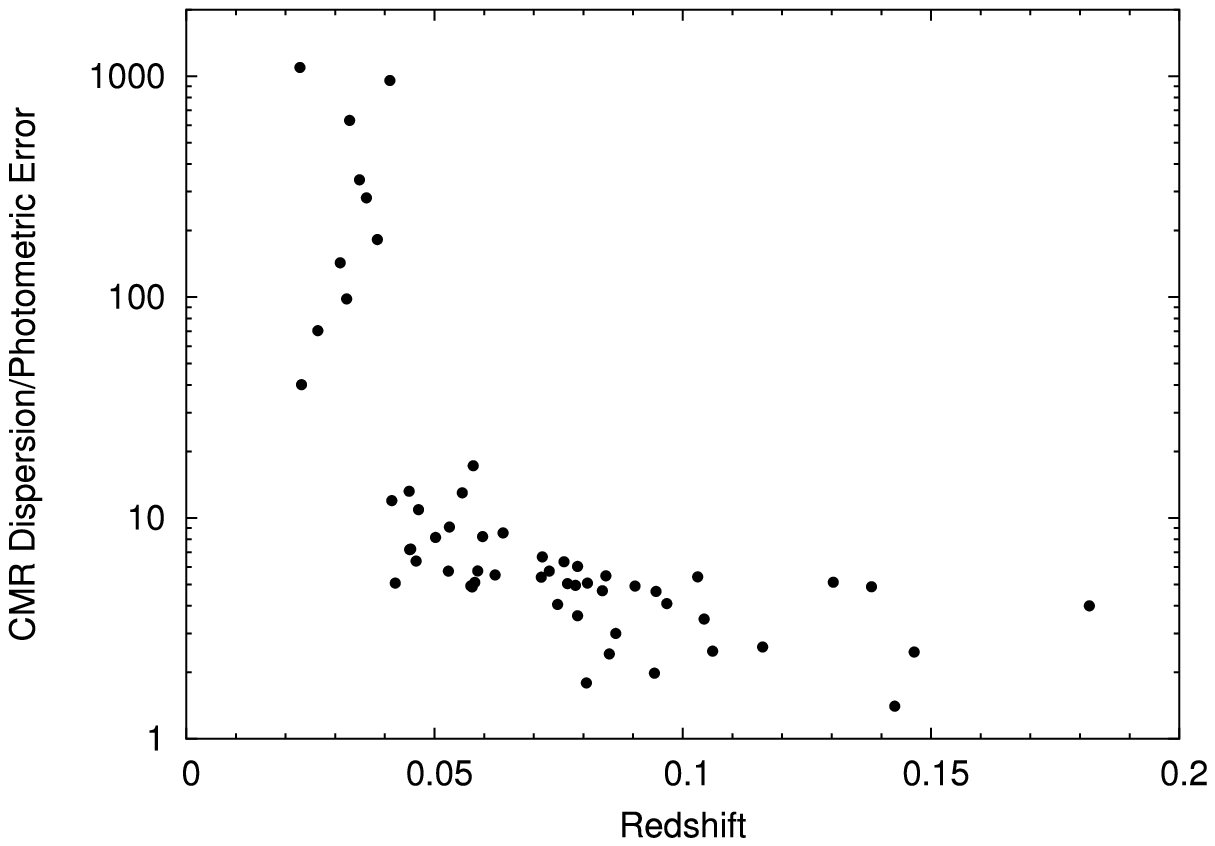}
\caption{}
\end{figure}
\begin{figure}
\figurenum{3}
\plotone{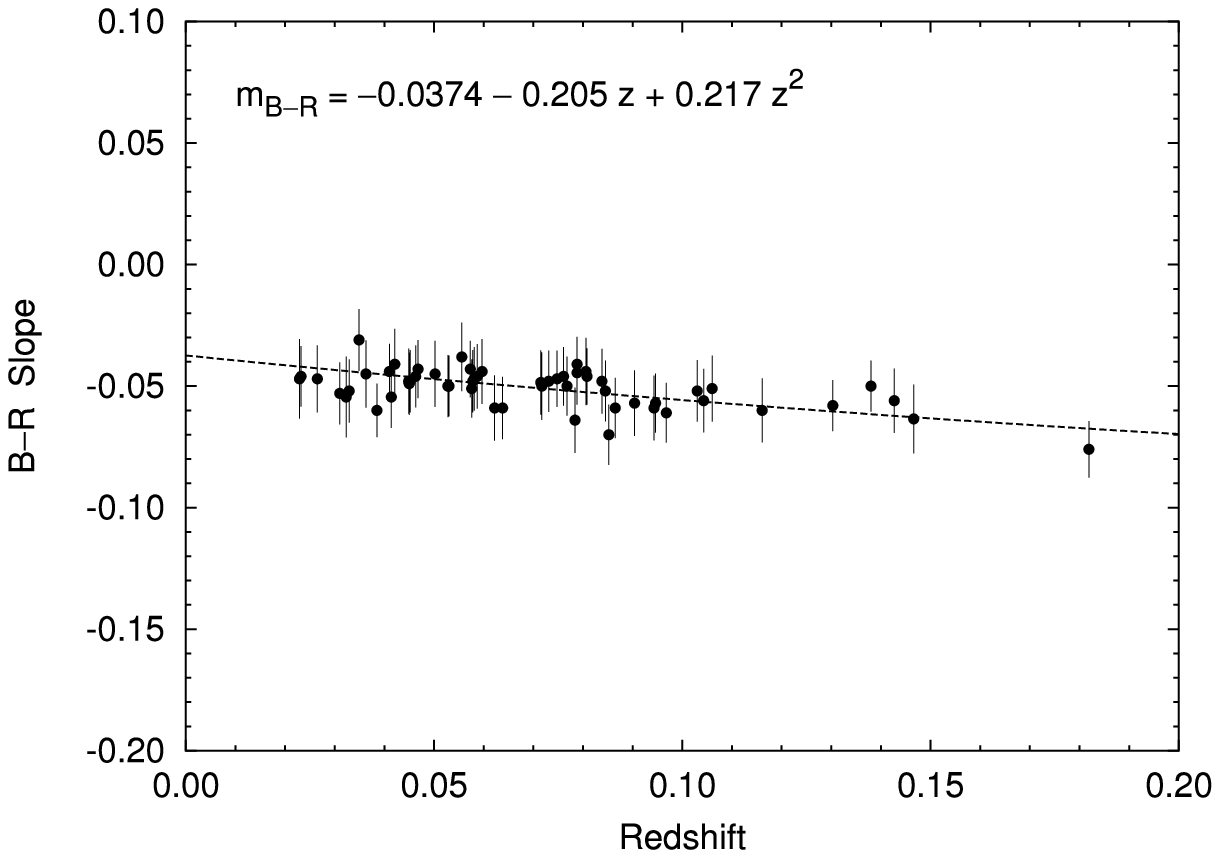}
\caption{}
\end{figure}
\begin{figure}
\figurenum{4}
\plotone{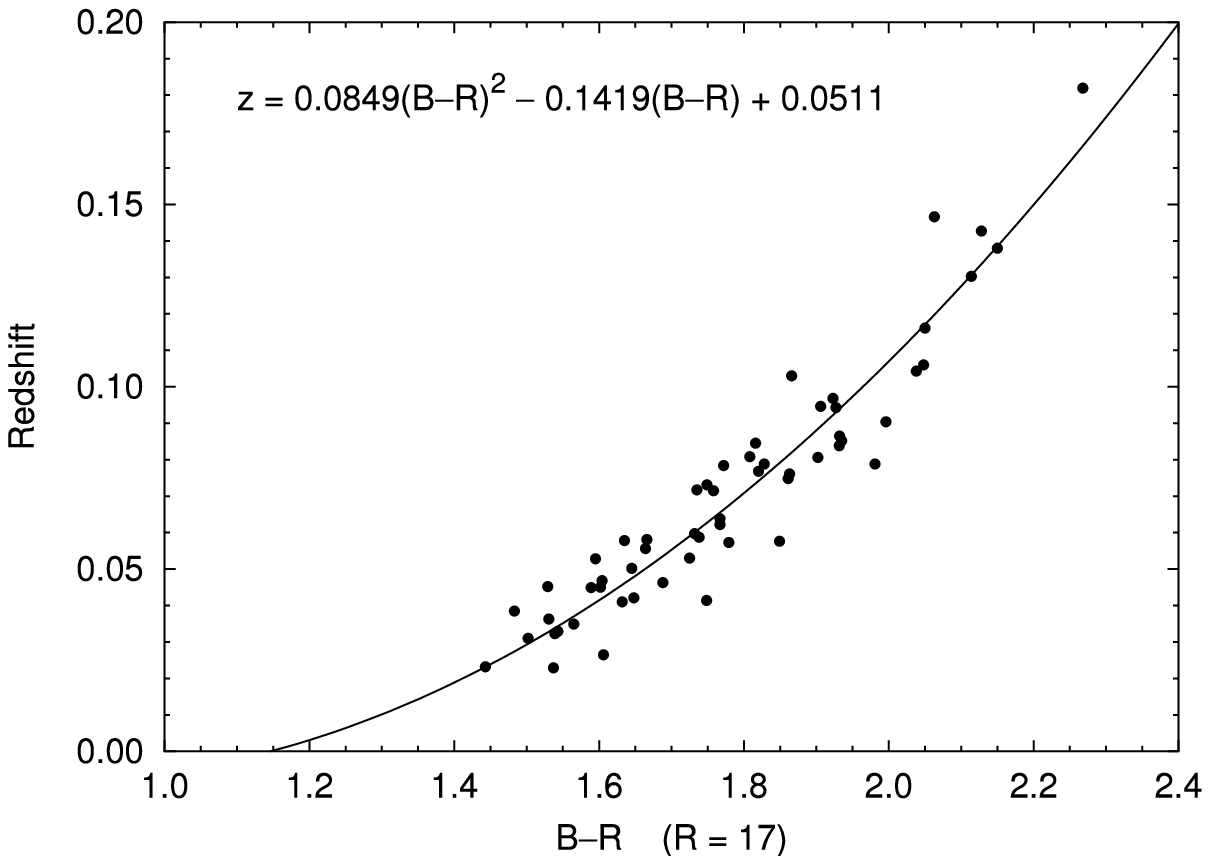}
\caption{}
\end{figure}

\clearpage

\begin{deluxetable}{cccccc}
\tablecaption{Cluster Sample and CMR Fits}
\tablehead{
\colhead{Cluster} &\multicolumn{1}{c}{Redshift($z$)}     
 &\colhead{Slope (b)}&\colhead{Zero Point (a)} &\colhead{$\sigma_{CMR}$}&
 \colhead{$(B-R)_{R=17}$}}

\startdata
A21    &  0.0946  &  -0.057 $\pm$ 0.012  &  2.88 $\pm$ 0.20  &   0.080   &  1.91   \\
A84    &  0.1030  &  -0.052 $\pm$ 0.013  &  2.75 $\pm$ 0.21  &   0.097   &  1.87   \\
A85    &  0.0556  &  -0.038 $\pm$ 0.014  &  2.31 $\pm$ 0.22  &   0.113   &  1.66   \\
A98    &  0.1043  &  -0.056 $\pm$ 0.013  &  2.99 $\pm$ 0.22  &   0.077   &  2.04   \\
A154   &  0.0638  &  -0.059 $\pm$ 0.013  &  2.77 $\pm$ 0.22  &   0.115   &  1.77   \\
A168   &  0.0452  &  -0.048 $\pm$ 0.013  &  2.34 $\pm$ 0.20  &   0.042   &  1.53   \\
A260   &  0.0363  &  -0.045 $\pm$ 0.014  &  2.30 $\pm$ 0.22  &   0.093   &  1.53   \\
A399   &  0.0715  &  -0.048 $\pm$ 0.013  &  2.58 $\pm$ 0.22  &   0.088   &  1.76   \\
A401   &  0.0748  &  -0.047 $\pm$ 0.012  &  2.66 $\pm$ 0.19  &   0.062   &  1.86   \\
A407   &  0.0463  &  -0.046 $\pm$ 0.013  &  2.47 $\pm$ 0.21  &   0.072   &  1.69   \\
A415   &  0.0788  &  -0.044 $\pm$ 0.013  &  2.74 $\pm$ 0.23  &   0.055   &  1.98   \\
A496   &  0.0329  &  -0.052 $\pm$ 0.013  &  2.43 $\pm$ 0.22  &   0.063   &  1.54   \\
A514   &  0.0731  &  -0.048 $\pm$ 0.013  &  2.56 $\pm$ 0.21  &   0.075   &  1.75   \\
A629   &  0.1380  &  -0.050 $\pm$ 0.010  &  3.00 $\pm$ 0.19  &   0.112   &  2.15   \\
A634   &  0.0265  &  -0.047 $\pm$ 0.014  &  2.40 $\pm$ 0.22  &   0.028   &  1.61   \\
A646   &  0.1303  &  -0.058 $\pm$ 0.010  &  3.10 $\pm$ 0.18  &   0.130   &  2.11   \\
A665   &  0.1819  &  -0.076 $\pm$ 0.012  &  3.56 $\pm$ 0.21  &   0.145   &  2.27   \\
A671   &  0.0502  &  -0.045 $\pm$ 0.014  &  2.41 $\pm$ 0.22  &   0.070   &  1.64   \\
A690   &  0.0788  &  -0.041 $\pm$ 0.011  &  2.52 $\pm$ 0.19  &   0.091   &  1.83   \\
A779   &  0.0229  &  -0.047 $\pm$ 0.016  &  2.34 $\pm$ 0.24  &   0.109   &  1.54   \\
A957   &  0.0450  &  -0.049 $\pm$ 0.013  &  2.44 $\pm$ 0.21  &   0.061   &  1.60   \\
A999   &  0.0323  &  -0.054 $\pm$ 0.017  &  2.46 $\pm$ 0.24  &   0.085   &  1.54   \\
A1142  &  0.0349  &  -0.031 $\pm$ 0.013  &  2.09 $\pm$ 0.22  &   0.034   &  1.56   \\
A1213  &  0.0468  &  -0.043 $\pm$ 0.012  &  2.34 $\pm$ 0.19  &   0.073   &  1.60   \\
A1291  &  0.0530  &  -0.050 $\pm$ 0.012  &  2.58 $\pm$ 0.21  &   0.077   &  1.72   \\
A1413  &  0.1427  &  -0.056 $\pm$ 0.013  &  3.08 $\pm$ 0.23  &   0.076   &  2.13   \\
A1569  &  0.0784  &  -0.064 $\pm$ 0.013  &  2.86 $\pm$ 0.22  &   0.067   &  1.77   \\
A1650  &  0.0845  &  -0.052 $\pm$ 0.012  &  2.70 $\pm$ 0.22  &   0.084   &  1.82   \\
A1656  &  0.0232  &  -0.046 $\pm$ 0.012  &  2.22 $\pm$ 0.21  &   0.056   &  1.44   \\
A1775  &  0.0717  &  -0.050 $\pm$ 0.014  &  2.58 $\pm$ 0.24  &   0.071   &  1.74   \\
A1795  &  0.0622  &  -0.059 $\pm$ 0.013  &  2.77 $\pm$ 0.23  &   0.067   &  1.77   \\
A1913  &  0.0528  &  -0.050 $\pm$ 0.013  &  2.44 $\pm$ 0.21  &   0.047   &  1.60   \\
A1983  &  0.0449  &  -0.048 $\pm$ 0.013  &  2.40 $\pm$ 0.22  &   0.062   &  1.59   \\
A2022  &  0.0578  &  -0.048 $\pm$ 0.013  &  2.45 $\pm$ 0.22  &   0.070   &  1.64   \\
A2029  &  0.0768  &  -0.050 $\pm$ 0.012  &  2.67 $\pm$ 0.20  &   0.080   &  1.82   \\
A2152  &  0.0410  &  -0.044 $\pm$ 0.011  &  2.38 $\pm$ 0.20  &   0.126   &  1.63   \\
A2244  &  0.0968  &  -0.061 $\pm$ 0.012  &  2.96 $\pm$ 0.21  &   0.073   &  1.92   \\
A2247  &  0.0385  &  -0.060 $\pm$ 0.011  &  2.50 $\pm$ 0.18  &   0.032   &  1.48   \\
A2255  &  0.0808  &  -0.046 $\pm$ 0.012  &  2.59 $\pm$ 0.20  &   0.073   &  1.81   \\
A2256  &  0.0581  &  -0.047 $\pm$ 0.013  &  2.46 $\pm$ 0.22  &   0.050   &  1.67   \\
A2271  &  0.0576  &  -0.051 $\pm$ 0.012  &  2.72 $\pm$ 0.20  &   0.059   &  1.85   \\
A2328  &  0.1466  &  -0.063 $\pm$ 0.014  &  3.14 $\pm$ 0.23  &   0.130   &  2.06   \\
A2356  &  0.1161  &  -0.060 $\pm$ 0.013  &  3.07 $\pm$ 0.24  &   0.061   &  2.05   \\
A2384  &  0.0943  &  -0.059 $\pm$ 0.013  &  2.93 $\pm$ 0.23  &   0.046   &  1.93   \\
A2399  &  0.0587  &  -0.046 $\pm$ 0.013  &  2.52 $\pm$ 0.22  &   0.070   &  1.74   \\
A2410  &  0.0806  &  -0.044 $\pm$ 0.014  &  2.65 $\pm$ 0.24  &   0.031   &  1.90   \\
A2415  &  0.0597  &  -0.044 $\pm$ 0.013  &  2.48 $\pm$ 0.23  &   0.089   &  1.73   \\
A2420  &  0.0838  &  -0.048 $\pm$ 0.013  &  2.75 $\pm$ 0.23  &   0.078   &  1.93   \\
A2440  &  0.0904  &  -0.057 $\pm$ 0.013  &  2.96 $\pm$ 0.23  &   0.088   &  2.00   \\
A2554  &  0.1060  &  -0.051 $\pm$ 0.014  &  2.92 $\pm$ 0.23  &   0.052   &  2.05   \\
A2556  &  0.0865  &  -0.059 $\pm$ 0.012  &  2.94 $\pm$ 0.21  &   0.049   &  1.93   \\
A2593  &  0.0421  &  -0.041 $\pm$ 0.014  &  2.34 $\pm$ 0.23  &   0.048   &  1.65   \\
A2597  &  0.0852  &  -0.070 $\pm$ 0.012  &  3.12 $\pm$ 0.24  &   0.037   &  1.94   \\
A2626  &  0.0573  &  -0.043 $\pm$ 0.012  &  2.51 $\pm$ 0.20  &   0.059   &  1.78   \\
A2634  &  0.0310  &  -0.053 $\pm$ 0.013  &  2.40 $\pm$ 0.20  &   0.047   &  1.50   \\
A2657  &  0.0414  &  -0.054 $\pm$ 0.013  &  2.68 $\pm$ 0.21  &   0.077   &  1.75   \\
A2670  &  0.0761  &  -0.046 $\pm$ 0.012  &  2.64 $\pm$ 0.20  &   0.094   &  1.86   \\
\enddata
\end{deluxetable}

\end{document}